\documentclass[twocolumn,showpacs,preprintnumbers,amsmath,amssymb,showkeys]{revtex4}

\usepackage{graphicx}
\usepackage{dcolumn}
\usepackage{bm}

\begin{document}

\preprint{APS/123-QED}

\title{Selective decay from a candidate of the $\sigma$-bond linear-chain state in $^{14}$C}

\author{J. Li$^1$}
\author{Y.L. Ye$^1$}
 \email{yeyl@pku.edu.cn}
\author{Z.H. Li$^1$}
\author{C.J. Lin$^2$}
\author{Q.T. Li$^1$}
\author{Y.C. Ge$^1$}
\author{J.L. Lou$^1$}
\author{Z.Y. Tian$^1$}
\author{W. Jiang$^1$}
\author{Z.H. Yang$^3$}
\author{J. Feng$^1$}
\author{P.J. Li$^1$}
\author{J. Chen$^1$}
\author{Q. Liu$^1$}
\author{H.L. Zang$^1$}
\author{B. Yang$^1$}
\author{Y. Zhang$^1$}
\author{Z.Q. Chen$^1$}
\author{Y. Liu$^1$}
\author{X.H. Sun$^1$}
\author{J. Ma$^1$}
\author{H.M. Jia$^2$}
\author{X.X. Xu$^2$}
\author{L. Yang$^2$}
\author{N.R. Ma$^2$}
\author{L.J. Sun$^2$}
\affiliation{$^1$School of Physics and State Key Laboratory of Nuclear Physics and Technology, Peking University, Beijing, 100871, China\\
$^2$China Institute of Atomic Energy,Beijing, 102413, China\\
$^3$RIKEN Nishina Center, 2-1 Hirosawa, Wako, Saitama 351-0198, Japan}



\date{\today}

\begin{abstract}
A cluster-transfer experiment $^9$Be($^9$Be,$^{14}$C$^*\rightarrow\alpha$+$^{10}$Be)$\alpha$
was carried out using an incident beam energy of 45 MeV. This reaction channel has a
large $Q$-value that favors populating the high-lying states in $^{14}$C and separating various reaction channels. A number of resonant states are reconstructed from the forward emitting $^{10}$Be + $\alpha$ fragments with respect to three sets of well discriminated final states in $^{10}$Be, most of which agree with the previous observations. A state at 22.5(1) MeV in $^{14}$C is found to decay predominantly into the states
around 6 MeV in $^{10}$Be daughter nucleus, in line with the unique property of the predicted band head of the
$\sigma$-bond linear-chain molecular states. A new state at 23.5(1) MeV is identified which decays strongly into the first excited state of $^{10}$Be.
\end{abstract}

\maketitle

$Introduction$. As one of the fundamental aspects in nuclear structure, a variety of cluster configurations is
known to exist in light stable nuclei \cite{vonOertzen200643,Horiuchi01012012}. For neutron-excess unstable nuclei, the presence of valence neutrons may help to stabilize the system, similar to the case of molecules \cite{vonOertzen200643,Horiuchi01012012}. Examples studied include neutron-excess beryllium isotopes which possess abundant molecular structure due to various valance-neutron configurations built on a well established $\alpha+\alpha$ rotor
of $^8$Be \cite{Zhyangprl,Zhyangprc,Zhyangsc,Lyu}. It is natural to extend such
ideas to carbon isotopes, which may contain three tightly bound $\alpha$ clusters\cite{vonOertzen200643,Itagaki}. Both triangle and linear-chain clustering states were predicted, based on effective interaction between the $\alpha$ clusters \cite{vonOertzen200643,Wikinson}.

A well established triple-$\alpha$ cluster state, the so-called
Hoyle state, was found in $^{12}$C at 7.65 MeV excitation level. This is featured by an $\alpha$-particle Bose-Einstein (BES) condensation \cite{vonOertzen200643}. Calculations using the molecular-orbital approach revealed instability of the linear-chain arrangement in $^{12}$C, but in the mean time it predicted this structure in $^{14}$C or $^{16}$C stabilized by the valence neutrons which form the $\pi$-orbital or $\sigma$-orbital molecular bond \cite{Itagaki}. The linear-chain configuration in $^{14}$C was also suggested by the antisymmetrized molecular dynamics (AMD) calculations \cite{Suhara}, which did not assume preformed cluster degree of freedom. Recently Baba $et$ $al.$, using an improved AMD code, made comprehensive calculations on $^{14}$C, which distinguishes the triangle-like structure and the $\pi$-bond and $\sigma$-bond linear-chain structures by the rotational band and decay pattern \cite{Baba}. The $\sigma$-bond chain state is predicted to possess exotic properties such as the longest chain distribution corresponding to the largest moment of inertia, the appearance at excitation energies beyond 22 MeV and most dramatically the selective decay path collecting to the $\sim$6 MeV states of the daughter nucleus $^{10}$Be \cite{Baba}. Considering the structural-link (overlap) between the mother and daughter nuclei, the latter property is expected since $\sim$6 MeV states in $^{10}$Be possess chain-like cluster structures as well \cite{Oertzen,vonOertzen200643}. The selective decay property is important for designing an experiment to search for the $\sigma$-bond linear-chain state.

So far quite intensive measurements have been carried out to study the cluster structure in $^{14}$C.
In an earlier $^7$Li($^9$Be,$^{14}$C$^*\rightarrow$$^{10}$Be$+\alpha$)$^2$H cluster-transfer experiment\cite{Soic}, the reaction $Q$-values (released energy) were applied to distinguish the ground state (g.s.), first excited state (3.4 MeV, $2^+$) and a group of excited states close to 6 MeV ($2^+, 1^-, 0^+, 2^-$) in the final decay-fragment $^{10}$Be. It was noticed that the states at 22.4(1) MeV and 24.0(3) MeV in $^{14}$C decay primarily into $\sim$6 MeV states of $^{10}$Be. The disadvantage of this earlier measurements is the large background below the $Q$-value peaks due to the employed reaction channel and the poor energy resolution of the detectors, which prohibited a clear identification of the reaction-decay process. A number of neutron transfer or inelastic excitation experiments were performed that populate the high-lying excited states in $^{14}$C, but are mostly insensitive to the final decay path collecting into $\sim$6 MeV states in $^{10}$Be, as summarised in Refs.\cite{Price,Haigh}. Some possible linear-chain states with $\pi$-bond feature \cite{Freer,Fritsch} were reported from the latest $\alpha$ resonant scattering experiments, however the decay to the $\sim$6 MeV states in $^{10}$Be were not resolved.

$Experiment$. We present here a measurement of the $\alpha$ decay from resonant states in $^{14}$C populated by the reaction $^9$Be($^9$Be,$^{14}$C$^*\rightarrow^{10}$Be+$\alpha$)$\alpha$.
This reaction channel has a very large $Q$-value so that high-lying excited states
can be easily populated and the decay-path be clearly separated. Preliminary results with partial data analysis were briefly reported in Ref.\cite{Tian}. The main focus of this article is the special high-lying excited states in $^{14}$C which decay predominantly into the $\sim$6 MeV states of $^{10}$Be, in line with the predicted property of the $\sigma$-bond linear-chain state \cite{Baba}.

The experiment was carried out at the HI-13 tandem accelerator at China Institute
of Atomic Energy (CIAE). A 166 $\rm {\mu g/cm^2}$ thick self-supporting $^9$Be target
was bombarded by a 45 MeV $^9$Be beam with an intensity of about 7 enA. The
charged fragments produced in the reactions were detected by six sets
of particle telescopes, namely U0,D0,U1,D1,U2,D2. The forward telescopes (U0 and D0) were located at
140 mm from the target and centered at $\pm 23^{\circ}$ relative to the beam direction.
Each of them consisted of two layers of double-sided silicon strip detector (DSSD) and one layer of large-size silicon detector (SSD), with
thicknesses of 50 $\mu$m (W1-60), 500 $\mu$m (BB7-500) and 1500 $\mu m$ (MSX40-1500), respectively.
The active area is 50 mm $\times$ 50 mm for W1-60 and 64 mm $\times$ 64 mm for
BB7-500 and MSX40-1500. Each side is divided into 16 or 32 strips for W1-60 or BB7-500, respectively. U1 and D1 were located at 116 mm from the
target and centered at $\pm 60 ^{\circ}$, while U2 and D2 at 114 mm and $\pm 109 ^{\circ}$.
U1, D1, U2 and D2 were consisted of identical elements: one W1-60 and one MSX40-1500.

Energy calibration of the six telescopes was realized by using $\alpha$-particle sources. In addition, the elastic scattering of $^9$Be from a $^{197}$Au target was also employed to calibrate the forward U0 and D0 telescopes which sustain broader energy ranges. The energy match for all silicon strips in one DSSD was achieved by the uniform calibration method described in Ref.\cite{RQiao}. The typical energy resolutions of the silicon detectors were less than 1.0\% for 5.49 MeV $\alpha$ particles. The position resolutions are determined by the respective strip widths.

The strip detector has a timing performance sufficient to be used to rule out most of the accidentally coincident signals. This is important since the strips close to the beam direction may record particles at a rate higher than $\sim 10^4$Hz. The beryllium and helium isotopes were
identified unambiguously by the standard energy loss versus residual energy ($\Delta E-E$) method. Reaction $Q$-value is a useful
quantity to select the reaction channel, since it is defined by the masses of the particles involved.
The largest possible $Q$-value has an obvious advantage of discerning contaminations from
other possible mass combinations.

\begin{figure}
  \centering
  \includegraphics[width=8cm]{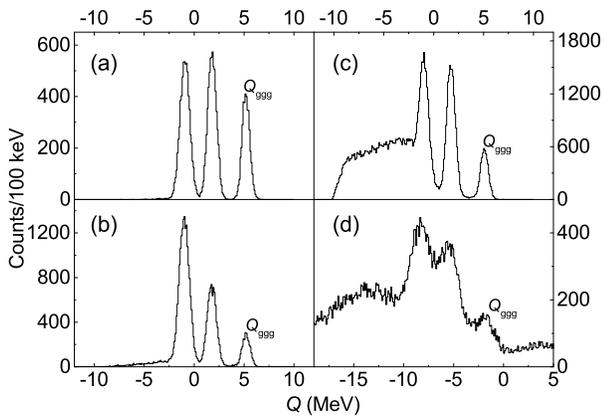}
  \caption{$Q$-value spectrum from present experiment for different data sets with: (a) identified
  $^{10}$Be and identified $\alpha$ in U0/D0 telescopes; (b) identified $^{10}$Be and unidentified
  $\alpha$; (c) identified $\alpha$ and unidentified $^{10}$Be. Spectrum (d) was obtained from previous
   $^7$Li($^9$Be,$\alpha^{10}$Be)$^2$H experiment \cite{Soic}. }\label{qvals}
\end{figure}

$Analysis$. Owing to the two deeply-bound $\alpha$-particles in the exit channel, the designed transfer-decay reaction $^9$Be($^9$Be,$^{14}\rm{C^*}\rightarrow\alpha +^{10}\rm{Be}$)$\alpha$ has a very large $Q$-value of 17.25 MeV for the intermediate population of $^{14}$C and also relatively large $Q$-value of 5.24 MeV for all final nucleus at their ground state (g.s.), namely $Q_{\rm ggg}$. The only contamination channel with the same final mass combination is the reaction $^9$Be($^9$Be,$^{10}$Be)$^8$Be. The latter can be easily removed by requiring a detectable $\alpha$-particle at the backward angle in coincidence with
the forward $^{10}$Be+$\alpha$ fragments taken by U0 and D0 telescopes. This coincidence is
kinematically characteristic of the former reaction-decay channel via $^{14}$C. From these triple-coincidence events
the reaction $Q$-value can be deduced according to
\begin{eqnarray}
Q&&= E_{tot}-E_{beam}\nonumber\\
&&= E_{^{10}\text{Be}}+E_{\alpha 1}+E_{\alpha 2}-E_{\text{beam}}\\
&&= M_{^{10}\text{Be}}+M_{\alpha 1}+M_{\alpha 2}-2 \times M_{^{9}\text{Be}} \nonumber
\label{Qvalequation}
\end{eqnarray}
Fig.\ref{qvals}(a) represents the experimental $Q$-value spectrum, from which
we see a very clean $Q_{ggg}$ peak together with another two clean peaks at
lower $Q$-values corresponding exactly to the first excited (3.4 MeV, $2^+$) and
$\sim$6 MeV states in $^{10}$Be \cite{Soic}. This background is largely reduced here thanks to
the selection of the reaction channel and the rejection of accidental coincidences by using
the fine timing information from the strips together with their good energy resolution.
This background free $Q$-value spectrum depicts the key advantage of the present experiment compared to the
previous ones when being applied to trace the decay path from $^{14}$C resonances.
In order to increase the event statistics, unidentified particles were also used to reconstruct the resonant states.
We may use two detected particles at the exit channel and deduce the third one based on energy-momentum conservation. The deduced energy-momentum were checked relative to the detected one assuming the targeted reaction process. Three sets of data samples are used in the
present analysis, all with $\alpha$-particles coincidentally detected at backward angles:

\begin{enumerate}
  \item Both identified $^{10}$Be and identified $\alpha$ at forward angles detected by U0/D0.
  The corresponding $Q$-value spectrum is shown in Fig.\ref{qvals}(a).
  \item Identified $^{10}$Be and unidentified $\alpha$ detected by U0/D0, with the latter being stopped
  in the first layer of the telescope due to an energy smaller than 8 MeV. The corresponding $Q$-value spectrum is presented in Fig.\ref{qvals}(b).
  \item Identified $\alpha$ and unidentified $^{10}$Be detected by U0/D0, with
   the latter being stopped in the first layer of the telescope for an energy between 8
  and 24 MeV. The corresponding $Q$-value spectrum is displayed in Fig.\ref{qvals}(c).

\end{enumerate}

In Fig.\ref{qvals} increasing background distribution below the $Q$-value peaks,
from (a) to (c), is shown. For selecting events that decay into the g.s. and first excited state (3.4 MeV) of $^{10}$Be
all three data sets can be used. For decaying into $\sim$6 MeV states, only data set 1 is a good choice. For comparison, the
$Q$-value spectrum measured in a previous cluster-transfer experiment is also presented in Fig.\ref{qvals}(d). This $Q$-value spectrum has
much higher background and also worse energy resolution.


\begin{figure}
  \centering
  \includegraphics[width=8cm]{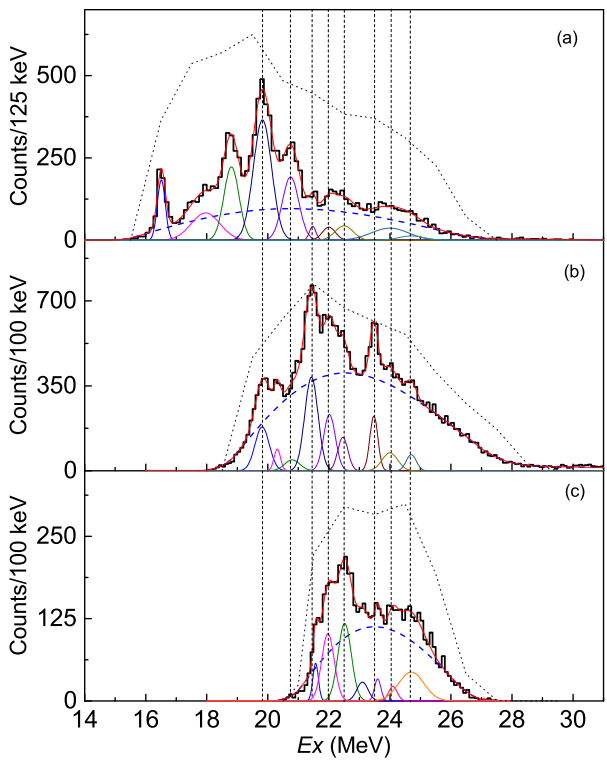}
  \caption{$^{14}$C excitation energy spectrum reconstructed from forward moving $^{10}$Be + $\alpha$ fragments,
  gated on the $Q$-value peaks for events decaying to the g.s (a), first excited state (3.4 MeV, 2$^+$) (b) and states around 6 MeV excitation (c), in
  $^{10}$Be. For (a) and (b) all three data sets are processed whereas for (c) only the data set 1 is used.
  Each spectrum is fitted by the sum (red solid line) of a smooth background (blue dashed line) \cite{Curtis} plus several Gaussian peaks. Black dotted lines illustrate the relative detection efficiency as a function of excitation energy. The vertical dotted lines are plotted to guide the eyes to the corresponding peaks. }  \label{exstates}
\end{figure}

By gating on the $Q$-value peaks corresponding to different final states in $^{10}$Be, the
relative energy of $^{14}$C can be reconstructed from the momenta of the
two forward moving $^{10}$Be + $\alpha$ fragments, with the invariant mass method \cite{Cao2012}. The corresponding $^{14}$C excitation energy
spectra are shown in Fig.\ref{exstates}. Peaks at 16.5, 19.8, 20.8, 21.4 and 22.5 MeV are observed in the present work and were also observed in previous experiment \cite{Soic,Price}, confirming the validity of the present measurement and reconstruction. In addition a new state at 23.5(1) MeV is clearly identified in the present measurement which decays strongly into the first excited state (3.4 MeV, $2^+$) in $^{10}$Be.

Monte Carlo simulations were performed to estimate the relative detection efficiencies with respect to three sets of final states in $^{10}$Be as illustrated in Fig.\ref{exstates}. These simulations include the details of detector geometries and their energy thresholds.
The angular distribution of the excited $^{14}$C is assumed to obey an exponential form whilst the emission of the decay products is treated isotropic in the center of mass (c.m.) system \cite{Price}. The systematic error associated with the simulation is estimated to be around 10\%, including the lack of spin information and the exact angular distribution of the highly excited $^{14}$C. The simulations also give energy resolutions of about 100$\sim$200 KeV for the energy range concerned.

\begin{table*}[t]
  \caption{Summary of the excited states populated in $^{14}$C and decaying to $\alpha$-cluster and $^{10}$Be in its ground, 2$^+$ and $\sim$6 MeV states. Those in square brackets represent tentative identifications.}\label{exstates_table}
  \begin{ruledtabular}
  \begin{tabular}{ccccccccc}
    \multicolumn{3}{c}{This work} & \multicolumn{3}{c}{$^7$Li($^9$Be,$\alpha^{10}$Be)$\alpha$\cite{Soic}} & \multicolumn{3}{c}{$^{14}$C($^{14}$C,$\alpha$$^{10}$Be)$^{14}$C\cite{Price}} \\
    \cline{1-3} \cline{4-6} \cline{7-9}
    \rule{0pt}{3ex}
    $^{10}$Be$_{\rm{gs}}$ & $^{10}$Be(2$^+$) &
    $^{10}$Be($\sim6$ MeV) & $^{10}$Be$_{\rm{gs}}$ & $^{10}$Be(2$^+$) &
    $^{10}$Be($\sim6$ MeV) & $^{10}$Be$_{\rm{gs}}$ & $^{10}$Be(2$^+$) &
    $^{10}$Be($\sim6$ MeV)\\
    \hline
    16.5(1) &  &  & 16.4(1) & & & 16.4(1) & & \\
     & & & & & & 17.3(1) & 17.3(1) & \\
    17.9(1) &  &  & & & & [18.1] & & \\
    18.8(1) &  &  & 18.5(1) & 18.5(1) & & 18.6 & 18.4(1) & \\
      &  &  & & [19.1(1)] & & & [19.0(2)]& \\
    19.8(1) & 19.8(1) & & 19.8(1) & & & & 19.8(1) & \\
      & 20.3(1) & & & & & & & 20.4(1) \\
    20.8(1) & 20.8 & & 20.6(1) & & & & & 20.9(1) \\
    \relax[21.4]\footnotemark[1] & 21.4(1) & 21.6(3) & & 21.4(1) & &  & [21.6(2)] & \\
    \relax[22.0]\footnotemark[1] & 22.0(1) & 22.0(3) & & & & & & [21.9(1)]\\
    \relax[22.5]\footnotemark[1] & 22.5(1) & 22.5(3) & & & 22.4(3) & & & [22.5(1)] \\
      &  & 23.1(3) & & [23.2(1)] & & & & [23.1(2)] \\
      & 23.5(1) & 23.6(3) & & & & & & \\
    \relax[24.0]\footnotemark[1] & [24.0(1)] & [24.1(3)] & & & 24.0(3) & & & \\
    \relax[24.7]\footnotemark[1] & [24.7(1)] & [24.7(3)] & & & & & & \\
  \end{tabular}
  \end{ruledtabular}
  \begin{flushleft}
  \footnotemark[1]{These states are assigned by comparison with neighbouring decay path.}
  \end{flushleft}
\end{table*}

The effective counts (relative decay branching ratios) associated with each resonance can be deduced by
\begin{equation}\label{counts}
  N_{\epsilon}=N/\epsilon
\end{equation}
where $\epsilon$ is the scaled detection efficiency determined by the simulation, and $N$ the number of events under each identified resonance peak.

In order to extract the peak centroid and the number $N$, each spectrum in Fig.\ref{exstates} was fitted with a series of Gaussian peak shapes plus a smoothly varying background \cite{Haigh,Curtis}. The peak centroids were initially allocated based on the visual identification and the previous reports in the literature on the existence of resonant states. Then peak centroids as well as widths were varied to minimize the $\chi^2$ value. In the present work the optimum fitting gives a $\chi^2$ per degree of freedom of 0.6$\sim$0.7. The obtained peak widths incorporate the energy resolutions of the detection system and also the intrinsic physics widths of the resonances. Some very broad peaks appear in the fitting configuration in the region of the quasi-continuous spectrum, each of which may cover unresolved multi-peaks. The extracted resonances are listed in Tab.\ref{exstates_table}. Numbers in square brackets mean that the peaks are not clearly discerned and the present identifications are just tentative. The error of the peak centroid is about 100 keV, contributed primarily from the systematical errors in the energy calibration of the telescopes and in the estimation of the dead-layer thickness of the silicon layers, and also from the statistical error determined by the fitting code. This error is further extended when decaying into $\sim$6 MeV combined states in $^{10}$Be \cite{Soic}. In the table resonances observed from the previous cluster transfer \cite{Soic} and the inelastic excitation \cite{Price} experiments are also presented for comparison.

\begin{figure}
  \centering
  \includegraphics[width=6cm]{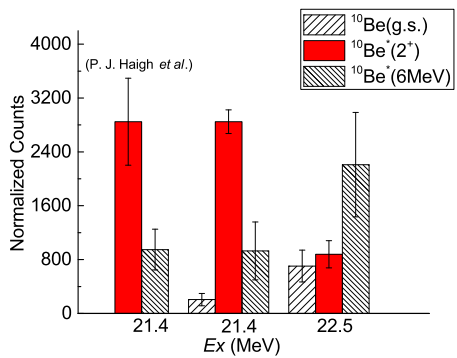}
  \caption{$^{14}$C $\rightarrow$ $^{10}$Be + $\alpha$ relative branching ratio for 21.4 and 22.5 MeV resonances in $^{14}$C with respect to three sets of final states in $^{10}$Be obtained from the present measurement. Results of the 21.4 MeV state taken from the previous 2-neutron transfer experiment \cite{Haigh} are also plotted for comparison. The error bars are statistical only.}\label{relabr}
\end{figure}

Fig.\ref{relabr} shows calculated relative decay branching ratios for the 21.4(1) and the 22.5(3) MeV resonances from the present work. Results from the 2-neutron transfer reaction \cite{Haigh} for the 21.4 MeV state is also plotted in the figure to demonstrate the consistency with the present work.

$Discussion$. Newly performed AMD calculations \cite{Baba}, with the Gogny D1S effective interaction, have successfully reproduced the threshold
and energies of the resonant states in $^{14}$C. The projection of the wave packet of the valence neutrons onto the single-particle states allows one to distinguish
the so-called $\pi$-bond or $\sigma$-bond molecular configurations. More importantly the decay properties of these resonances are studied comprehensively with respect to various final states in the $^{10}$Be daughter nucleus, which are obviously instructive for experimental investigations. The calculation shows that the $\sigma$-bond linear-chain molecular band (positive parity), with a $0^+$ band-head starting from 22.16 MeV, is situated appreciably above other triangle-like or $\pi$-bond chain states (see Table II and Figure 6 in Ref.\cite{Baba}). In addition, these $\sigma$-bond chain states decay almost exclusively into the $\sim$6 MeV states in $^{10}$Be, whereas other resonances in $^{14}$C decay mostly into rotational states associated with ground band in $^{10}$Be including its g.s. and the first excited state(3.4 MeV, $2^+$). This unique decay feature of the $\sigma$-bond chain state can be understood by the structure change in the daughter nucleus $^{10}$Be. As described in Ref.\cite{Oertzen1996,Oertzen}, the ground and $2_1^+$ states of $^{10}$Be possess primarily the single-particle structure whereas the four states around 6 MeV have typical two-center chain-type configurations. Therefore it is natural that the latter provides strong collection (wave-function overlap) with the typical cluster states in the mother nucleus $^{14}$C.

\begin{figure}
  \centering
  \includegraphics[width=6cm]{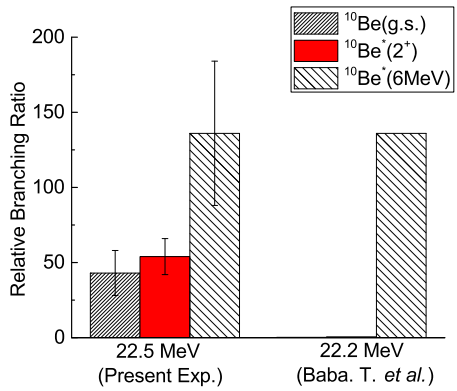}
  \caption{Comparison of the relative decay branching ratio obtained from the present experiment with the theoretical prediction \cite{Baba}.
  }\label{225MeV}
\end{figure}

From the present experiment, we found a resonance at 22.5(3) MeV in $^{14}$C which decays primarily into the $\sim$6 MeV states in $^{10}$Be as displayed qualitatively in Fig.\ref{exstates} and plotted quantitatively in Fig.\ref{225MeV}. This is consistent with a previous measurement for $^{14}$C excitation \cite{Soic}, in which a peak at 22.4(3) MeV was found to decay mainly to the $\sim$6 MeV states in $^{10}$Be although its selection on $Q$-value is questionable as discussed above. The observed state corresponds well to the predicted band-head at 22.2 MeV of the $\sigma$-bond linear-chain molecular band \cite{Baba}, based on the similar excitation energy and the unique decay feature as shown in Fig.\ref{225MeV}. Some minor differences exist between the prediction and observation. The former having a pure decay branching-ratio while the latter decaying also into two other low lying states of $^{10}$Be in small factions. One possible explantation is that the observed 22.5(3) MeV state also contains minor mixtures other than $\sigma$-bond linear-chain structure. One question that might be raised is what are the relative contributions from the four close-by states around 6 MeV in $^{10}$Be, namely $2^+$ state at 5.958 MeV, $1^-$ at 5.96 MeV, $0^+$ at 6.18 MeV and $2^-$ at 6.26 MeV. As a matter of fact, by adopting the observed 22.5 MeV resonance in $^{14}$C as the predicted $0^+$ band-head, as discussed above, the predominantly collected final state in $^{10}$Be should be its $0_2^+$ at 6.18 MeV. The reason is that this $0_2^+$ state possess a pure $\sigma$-bond configuration while the other three are characterized by $\pi$-orbit \cite{Oertzen1996,Oertzen} (decaying to $2^-$ state is also prohibited by parity conservation). This selection among the four close-by final states was also confirmed by the authors of Ref.\cite{Baba} via the private communications.

It is noted that, unlike the very selective collection between $\sigma$-bond linear-chain state in $^{14}$C and
the $\sim$6 MeV states in the $^{10}$Be daughter nucleus, the decay strengths for triangle-like and $\pi$-bond chain state feeding into
the g.s. and the first excited state of $^{10}$Be varies case by case as can be seen from Ref.\cite{Baba} and from the above Fig.\ref{relabr}.
More studies for these states are certainly needed to find correspondence between theoretical calculations and experimental observations .
\\

$Summary$. A cluster transfer experiment, $^9$Be($^9$Be,$^{14}$C$\rightarrow\alpha$+$^{10}$Be)$\alpha$, was carried out with a 45 MeV beam energy. This reaction channel has an extremely large $Q$-value in favor of the $^{14}$C excitation and the selection of the reaction-decay processes. A clean $Q$-value spectrum was obtained which allows one to trace the decay path related to the structures of the mother and daughter nuclei. A number of resonances in $^{14}$C are observed from the $\alpha$ decay channel. Most of these resonaces agree with the previously reported results\cite{Soic,Haigh}. It is found that the 22.5(1) MeV state in $^{14}$C decays predominantly into the $\sim$6 MeV states in $^{10}$Be, in line with the unique feature of the predicted $\sigma$-bond linear-chain state in $^{14}$C. The excitation energy of this cluster resonance consists also with the band head of the $\sigma$-bond linear-chain rotation which appears appreciably above other molecular rotational band. The present work, together with the latest theoretical calculation of Baba $et$ $al.$, encourages the searching for the most exotic $\sigma$-bond chain state with exclusive experimental signatures. A new state at 23.5 MeV is clearly identified from the present measurement which decays primarily into the first excited state of $^{10}$Be. Its properties still need to be investigated. The observations would be more conclusive if the spins of the high-lying resonances were determined from the measurement. This can be achieved in future experiments with detection setups more sensitive to the required angular corrections.

$Acknowledgement$. The authors would like to give special thanks to Dr. T. Baba and Dr. M. Kimura of the Hokkaido University for the active discussions and the theoretical support. We are grateful to the staff of the HI-13 tandem acceleartor for their excellent work in providing the beams. This work is supported by the 973 Program of China (No. 2013CB834402), the National Natural Science Foundation of China (No.11535004, No.11275011, No.11375017, No. 11275001)

\bibliography{c14-clustering-R1}

\end{document}